\documentclass[fleqn,prl,twocolumn]{revtex4}

\usepackage{amsmath,graphicx}
\usepackage{dcolumn}
\usepackage{verbatim}

\begin{document}

\title{Test of the theoretical hyperfine structure of \\
the molecular hydrogen ion at the 1-ppm level}

\author{Vladimir I. Korobov}
\affiliation{Bogoliubov Laboratory of Theoretical Physics, Joint Institute
for Nuclear Research, Dubna 141980, Russia}
\author{J.C.J.~Koelemeij}
\affiliation{LaserLaB and Department of Physics \& Astronomy, VU University, De Boelelaan 1081, 1081 HV Amsterdam, The Netherlands}
\author{L.~Hilico}
\author{J.-Ph.~Karr}
\affiliation{Laboratoire Kastler Brossel, UPMC-Sorbonne Universit\'es, CNRS, ENS-PSL Research University, Coll\`ege de France\\
4 place Jussieu, F-75005 Paris, France}
\affiliation{Universit\'e d'Evry-Val d'Essonne, Boulevard Fran\c cois Mitterrand, F-91000 Evry, France}

\begin{abstract}
We revisit the $m \alpha^6 (m/M)$ order corrections to the hyperfine splitting in the H$_2^+$ ion, and find a hitherto unrecognized second-order relativistic contribution associated with the vibrational motion of the nuclei. Inclusion of this correction term produces theoretical predictions which are in excellent agreement with experimental data  [K. B. Jefferts, Phys.\ Rev.\ Lett.\ \textbf{23}, 1476 (1969)], thereby concluding a nearly fifty years long theoretical quest to explain the experimental results within their 1-ppm error. The agreement between theory and experiment corroborates the proton structural properties as derived from the hyperfine structure of atomic hydrogen. Our work furthermore indicates that for future improvements, a full three-body evaluation of the $m \alpha^6 (m/M)$ correction term will be mandatory.
\end{abstract}

\maketitle

One of the key properties of H$_2^+$, the simplest molecule in nature, is the frequency of its hyperfine transitions, which include the molecular counterparts to the well-known 21-cm line in atomic hydrogen. Thirty of these hyperfine transitions were investigated back in 1969 in a pioneering radio-frequency (rf) spectroscopy experiment~\cite{Jeff69} with an accuracy of 1.5~kHz (or 1~ppm). Indirect information was later obtained, with lower accuracy, from the analysis of Rydberg states spectra in H$_2$~\cite{Fu92,Ost04}. Many theoretical efforts have since been devoted to explaining the hyperfine structure (hfs) of H$_2^+$. First calculations within the adiabatic approximation (see \textit{e.g.}~\cite{Luke69,McE78}) were in disagreement with experiments by slightly less than 1~MHz. The inclusion of nonadiabatic corrections~\cite{Babb92}, and a full three-body calculation of the leading-order hyperfine Hamiltonian~\cite{HFS06} allowed to reduce the discrepancy by about one order of magnitude. Recently, the theory was further improved by including relativistic corrections of order $m \alpha^6 (m/M)$~\cite{HFS09} but a discrepancy of $\sim 10$~kHz between the theoretical and experimental spin-flip transition frequencies still remained unexplained.

This situation may be considered discomforting for three main reasons. Firstly, H$_2^+$ serves as a benchmark system for high-accuracy theoretical models. Any unexplained discrepancy between theory and experiment may point at possible issues with theory (in this case relativistic quantum mechanics and quantum-electrodynamics), experiment, or both. Secondly, H$_2^+$ is assumed to play a key role in the formation of tri-hydrogen molecular ions in space, which are the seed of many interstellar chemical reaction chains~\cite{Oka92}. In contrast to H$_3^+$, interstellar H$_2^+$ has remained elusive, and its detection is one of the outstanding challenges in radio astronomy~\cite{Luke69,Penzias68,Jeff70,Shuter86,Varsh93}. A possible future radio-astronomical detection of H$_2^+$ hyperfine emission lines will likely trigger subsequent modeling of the cloud dynamics based on Doppler velocity profiles. This process (and even the identification of the hyperfine lines themselves) may be hampered by the above-mentioned hyperfine frequency discrepancy, which could be mistaken for a large Doppler shift $v/c \sim 10^{-5}$. Thirdly, there exists a possibility that the discrepancy could be at least partly explained by a significant deviation of the proton's structural parameters (Zemach radius and polarizability) from their current values, similar to the smaller electric charge radius observed in muonic hydrogen~\cite{Pohl10}. Resolving the discrepancy is therefore of importance to a broad physics community.

In this Letter, we point out a hitherto unrecognized deficiency in the evaluation of the relativistic correction to the $m \alpha^6 (m/M)$ contribution, which we solve \textit{ad hoc} by the inclusion of vibrational excitations within this correction term. With this correction included, our calculations are in excellent agreement with experimental data, for the first time within the experimental error, thereby concluding a theoretical quest which lasted for nearly half a century (see for example~\cite{Babb08} and references therein). Furthermore, the agreement between theory and experiment at the kHz level corroborates the proton nuclear properties derived from the experimentally measured hfs of atomic hydrogen~\cite{Shabaev}. At the same time, our findings suggest that the current theoretical framework should be extended with a rigorous treatment of the entire $m \alpha^6 (m/M)$ relativistic correction in the three-body framework in order to achieve higher precision in the future.

Our theoretical treatment of the H$_2^+$ hyperfine structure starts out from the nonrelativistic Hamiltonian of a three-body system, which may be written as follows (atomic units are used throughout):
\begin{equation}\label{H3body}
H =
   -\frac{1}{2\mu}\nabla^2_{r_1}
   -\frac{1}{2\mu}\nabla^2_{r_2}
   -\frac{1}{m_e}\nabla_{r_1}\cdot\nabla_{r_2}
   -\frac{Z}{r_1}-\frac{Z}{r_2}+\frac{Z^2}{r_{12}}\,,
\end{equation}
where $\mathbf{r}_1$ and $\mathbf{r}_2$ are position vectors of the electron with respect to protons labeled 1 and 2, $\mathbf{r}_{12}=\mathbf{r}_2-\mathbf{r}_1$ is a vector determining the relative position of the two protons, $Z=1$ is the nuclear charge, and $\mu=m_eM_p/(m_e+M_p)$ is the reduced mass given the electron mass, $m_e$, and proton mass, $M_p$.

\begin{figure}[t]
\begin{center}
\includegraphics[width=0.38\textwidth]{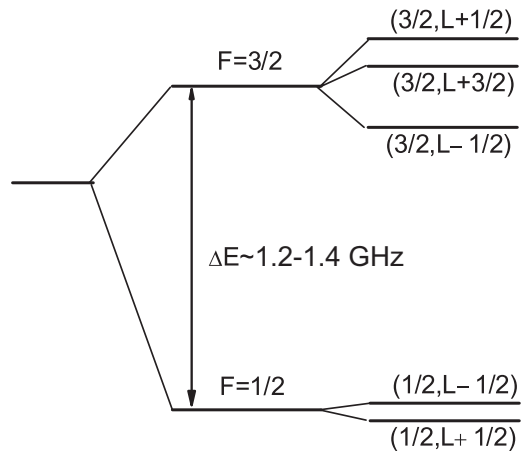}
\end{center}
\caption{Hyperfine structure of $\mbox{H}_2^+$ in a given vibrational quantum state $v$ and with orbital angular momentum quantum number $L=1$. Angular momentum quantum labels $F$ and $J$ are shown within parentheses.}\label{fig:HFS}
\end{figure}

The hyperfine interaction, which determines the splitting between various spin configuration sublevels within the same rovibrational state in the $\mbox{H}_2^+$ ion, is described by the effective Hamiltonian~\cite{HFS06}
\begin{equation}\label{effH_H2}
\begin{array}{@{}l}
\displaystyle H_{\rm eff} =
    b_F(\mathbf{I}\cdot\mathbf{s}_e)
   +c_e(\mathbf{L}\cdot\mathbf{s}_e)
   +c_I(\mathbf{L}\cdot\mathbf{I})
      +\frac{d_1}{(2L\!-\!1)(2L\!+\!3)}
\\[2mm]\displaystyle
\hspace{9mm}
\times \;   \biggl\{
          \frac{2}{3}\mathbf{L}^2(\mathbf{I}\cdot\mathbf{s}_e)
          -[(\mathbf{L}\cdot\mathbf{I})(\mathbf{L}\cdot\mathbf{s}_e)
           +\!(\mathbf{L}\cdot\mathbf{s}_e)(\mathbf{L}\cdot\mathbf{I})]
       \biggr\}
\\[3mm]\displaystyle
\hspace{9mm}
   +\frac{d_2}{(2L\!-\!1)(2L\!+\!3)}
       \left[
          \frac{1}{3}\mathbf{L}^2\mathbf{I}^2
          -\frac{1}{2}(\mathbf{L}\cdot\mathbf{I})
          -(\mathbf{L}\cdot\mathbf{I})^2
       \right].
\end{array}
\end{equation}
Following the notation of Ref.~\cite{HFS06}, $\mathbf{I}$ is the total nuclear spin, and $\mathbf{L}$ is the total orbital momentum. The coupling scheme of angular momenta is $\mathbf{F}=\mathbf{I}+\mathbf{s}_e$, $\mathbf{J}=\mathbf{L}+\mathbf{F}$. A schematic diagram indicating the hyperfine splitting between the states with rotational angular momentum quantum number $L=1$ is shown in Fig.~\ref{fig:HFS}.

The major coupling is the spin-spin electron-proton interaction (first term in Eq.~(\ref{effH_H2})) which determines the principal splitting between $F\!=\!1/2$ and $F\!=\!3/2$ states of $\sim$1.2-1.4~GHz. Thus, the main contribution to the theoretical uncertainty of the spin-flip transition frequencies stems from uncertainty of the coefficient $b_F$. In what follows we will concentrate on calculations of this quantity. The coefficients $c_e$, $c_I$, $d_1$, and $d_2$ of the Hamiltonian~(\ref{effH_H2}) were obtained in~\cite{HFS06} with sufficient accuracy to explain the smaller splitting of the $F\!=\!1/2$ and $F\!=\!3/2$ multiplets within the experimental uncertainty.

In \cite{HFS09} the following QED contributions to $b_F$ were taken into consideration: $(Z\alpha)^2E_F$~\cite{Breit}, $\alpha(Z\alpha)E_F$~\cite{Kroll,Schwin}, $\alpha(Z\alpha)^2\ln^2(Z\alpha)E_F$~\cite{Layzer}, as well as the proton finite size corrections: Zemach~\cite{Zemach,Karsh97}, pure recoil~\cite{BodYen88}, and nuclear polarizability~\cite{Fau02} contributions (see also~\cite{Shabaev,Carlson} for more details).
\[
E_F =\frac{16}{3} \alpha^2 \mu_p \frac{m_e}{M_p} \left[1+\frac{m_e}{M_p}\right]^{-3}
\]
is the Fermi energy for the hyperfine splitting in atomic hydrogen (H) \cite{SapYen,Kin96}, with $\mu_p$ the magnetic moment of the proton in nuclear magnetons. The work of Ref.~\cite{HFS09} allowed to reduce the discrepancy with measured spin-flip transitions $(F=3/2)\to(F=1/2)$ by one order of magnitude in comparison with the Breit-Pauli Hamiltonian approximation of Ref.~\cite{HFS06}, down to about 10~kHz. Here, we revisit the $(Z\alpha)^2E_F$ contribution. The other QED contributions considered in~\cite{HFS09} are proportional to the electronic wave function density $|\Psi(0)|^2$ at the location of each nucleus, and were obtained with sufficient precision.

We evaluate the interactions of order $(Z \alpha)^2 E_F$ within the framework of non-relativistic quantum electrodynamics (NRQED)~\cite{Kin96}. The NRQED spin-spin interaction at this order has two terms, which contribute to the hyperfine splitting. The first one is the effective Hamiltonian  (see~\cite{HFS09}, Eq.~(43)):
\begin{equation}\label{effH}
\begin{array}{@{}l}
\displaystyle
\Delta E_B = \left\langle H_s^{(6)} \right\rangle, \hspace{2mm} H_s^{(6)}
 = \alpha^4\frac{Z}{3m_e}\frac{\mu_p}{m_p}
   \left[
      -\frac{1}{4m_e^2}\Bigl\{p_e^2, \right.
\\[2mm] \displaystyle
        \left.   4\pi
         \left[
            \delta(\mathbf{r}_1)\!+\!\delta(\mathbf{r}_2)
         \right]\Bigr\}
      +\frac{Z}{2m_e}
      \left(
         \frac{1}{r_1^4}+\frac{1}{r_2^4}
         +\frac{2\mathbf{r}_1\mathbf{r}_2}{r_1^3r_2^3}
      \right)
   \right]
   \left(\mathbf{s}_e\!\cdot\mathbf{I}\right),
   \end{array}
\end{equation}
and the other is the second-order contribution:
\begin{equation}\label{eq:2order}
\Delta E_A = 2\alpha^4
   \left\langle
      H_B\left|Q(E_0-H)^{-1}Q\right|H_{ss}
   \right\rangle,
\end{equation}
where
\begin{equation}
\begin{array}{@{}l}
\displaystyle
H_B = -\frac{\mathbf{p}_e^4}{8m_e^3}
   +\frac{Z}{2m_e^2}\pi\left[\delta(\mathbf{r}_1)+\delta(\mathbf{r}_2)\right],
\\[2mm] \displaystyle
H_{ss} = \frac{8\pi}{3m_e}\mathbf{s}_e
   \left[\boldsymbol{\mu}_1\delta(\mathbf{r}_1)+\boldsymbol{\mu}_2\delta(\mathbf{r}_2)\right].
   \end{array}
\end{equation}
with $\boldsymbol{\mu}_a=(\mu_p/M_p)\mathbf{I}_a$ for $a=1,2$. In Ref.~\cite{HFS09} the $(Z \alpha)^2 E_F$ correction was evaluated within the adiabatic approximation, \textit{i.e.} it was computed for the two Coulomb center problem for a grid of values of the internuclear distance $r_{12}$ to obtain an ''effective'' potential curve, which was then averaged over the vibrational motion of heavy particles in the molecule. The Hamiltonian $H$ in the second-order term was thus approximated by the two-center Hamiltonian $H_0$,
\begin{equation}\label{H2center}
H_0 = \frac{p_e^2}{2m_e}-\frac{Z}{r_1}-\frac{Z}{r_2},
\end{equation}
where $r_1$ and $r_2$ are distances from clamped proton 1 and 2 to the electron, respectively. The major deficiency of the adiabatic approach is that for the second-order term, only the electron excitations are taken into consideration, whereas vibrational excitations are ignored. To solve this, we explicitly incorporate the vibrational excitations in our analysis by adding a third contribution
\begin{equation}\label{eq:2order3}
\begin{array}{@{}l}
\displaystyle \Delta E_C = 2\alpha^4 \left\{
\sum_{v'\ne v} \frac{\left\langle vL |H_B| v'L\right\rangle\left\langle v'L |H_{ss}| vL\right\rangle} {E_v-E_{v'}} \right.
\\[3mm] \displaystyle \hspace{11mm} \left. +
\int_{E_{1S}}^{\infty} \frac{\left\langle vL |H_B| E L\right\rangle\left\langle E L |H_{ss}| vL\right\rangle} {E_v-E} \; dE
\right\}.
\end{array}
\end{equation}
The first and second term correspond respectively to a summation over vibrational and continuum states within the ground electronic state only, and $E_{1S}$ stands for the energy of the dissociation limit of the $1s\sigma_g$ electronic ground state of H$_2^+$. For the contribution from excited $\sigma_g$ electronic states, we keep the result of the adiabatic calculation performed in~\cite{HFS09}, thereby ignoring the vibrational excitations associated with these states. The underlying assumption is that the correlation with vibrational states in excited electronic states is negligible due to large energy difference and small overlap of these vibrational wavefunctions with those in the $1s\sigma_g$ state.

Both bound and continuum wave functions of the H$_2^+$ molecular ion are required for a numerical evaluation of Eq.~(\ref{eq:2order3}). For bound ro-vibrational states of \emph{gerade} symmetry we use the variational method described in \cite{var00}. To get proper solutions for all $L\!=\!1$ states up to the last vibrational state $v=19$, basis sets with $N\!=\!2000$ functions for the states of $v=0,\dots,12$, and $N=3000$ functions for the states of $v=13,\dots,19$ are used. This approach enables us to obtain the nonrelativistic energies for these states ($E_v,E_v'$) with at least twelve significant digits, and to obtain the matrix elements appearing in Eq.~(\ref{eq:2order3}) with sufficient precision. For continuum states (with energy $E$), Born-Oppenheimer wave functions are generated from the $1s\sigma_g$ electronic curve.

\begin{table}[t]
\begin{center}
\begin{tabular}{@{\hspace{2mm}}c@{\hspace{5mm}}r@{\hspace{5mm}}r@{\hspace{5mm}}r@{\hspace{2mm}}}
\hline\hline
\vrule width0pt height10pt depth4pt
contribution & $v=0$~~&  $v=4$~~&  $v=8$~~ \\
\hline
\vrule width0pt height10pt depth4pt
$b_F$ \cite{HFS06}   &  922.9918     &  836.7835      &  775.2206     \\
$(Z\alpha)^2$        &   0.0663      &    0.0607      &    0.0569     \\
                     &  (0.0513)\!\! &   (0.0510)\!\! &   (0.0515)\!\!    \\
$\alpha(Z\alpha)$    & $-$0.0887     & $-$0.0804      & $-$0.0745     \\
$\alpha(Z\alpha)^2\ln^2(Z\alpha)$
                     & $-$0.0074     & $-$0.0067      & $-$0.0062     \\
$\Delta E_Z$         & $-$0.0369     & $-$0.0335      & $-$0.0310     \\
$\Delta E^p_R$       &    0.0054     &    0.0049      &    0.0045     \\
$\Delta E_{\rm pol}$ &    0.0013     &    0.0012      &    0.0011     \\
\hline
\vrule width0pt height10pt depth4pt
$b_F$(this work)     &  922.9318     &  836.7294 &  775.1714 \\
\hline\hline
\end{tabular}
\end{center}
\caption{Summary of the contributions to the spin-spin interaction coefficient $b_F$ (in MHz). To illustrate the contribution of Eq.~(\ref{eq:2order3}), which was hitherto neglected, the value of the $(Z\alpha)^2$ term taken from Ref.~\cite{HFS09} is shown within parentheses.}\label{tab:contrib}
\end{table}

\begin{table}[t]
\begin{center}
\begin{tabular}{@{\hspace{2mm}}c@{\hspace{5mm}}c@{\hspace{5mm}}c@{\hspace{5mm}}c@{\hspace{5mm}}l@{\hspace{2mm}}}
\hline\hline
\vrule width0pt height10pt depth4pt
 & \multicolumn{3}{c}{$L=1$} \\
\cline{2-4}
\vrule width0pt height10pt depth4pt
$v$ & \cite{HFS06} & \cite{HFS09} & this work & experiment \\
\hline
\vrule width0pt height10pt depth4pt
0   & 922.992    & 922.917   & 922.9318  & 922.940(20)~\cite{Fu92} \\

    &            &           &           & 923.16(21)~\cite{Ost04} \\
1   &  898.809   &  898.737  &  898.7507 & \\
2   &  876.454   &  876.385  &  876.3973 & \\
3   &  855.812   &  855.746  &  855.7570 & \\
4   &  836.784   &  836.720  &  836.7294 &  836.729  (1)\\
5   &  819.280   &  819.219  &  819.2272 &  819.227  (1)\\
6   &  803.227   &  803.167  &  803.1750 &  803.175  (1)\\
7   &  788.558   &  788.501  &  788.5079 &  788.508  (1)\\
8   &  775.221   &  775.166  &  775.1714 &  775.172  (1)\\
\hline\hline
\end{tabular}
\end{center}
\caption{Values of the spin-spin interaction coefficient $b_F$ in MHz (third column), and comparison with previously published theoretical values (first and second columns) and experimental ones  wherever available (last column). The results of Ref.~\cite{HFS06} contain no correction of order $(Z \alpha)^2 E_F$, whereas those of Ref.~\cite{HFS09} include the $(Z \alpha)^2 E_F$ correction but without the contribution by vibrational excitations. Experimental values for $v = 4-8$ are taken from~\cite{Jeff69}.}\label{comparison}
\end{table}

Table \ref{tab:contrib} illustrates the relative importance of the various contributions to the $b_F$ coefficient. Below the $(Z\alpha)^2E_F$ term we give within parentheses the previous value taken from \cite{HFS09}. Final values of $b_F$ for the vibrational states $v=0-8$ are given in Table~\ref{comparison} along with previously published theoretical values. We find that the contribution by vibrational excitations amounts to $23\%$ of the total $(Z \alpha)^2 E_F$ order correction for the ground ($v=0$) state, and decreases with increasing vibrational quantum number (\textit{e.g.} to $16\%$ for $v=4$ and $9\%$ for $v=8$). The dominant effect is that of the bound vibrational states (first term in Eq.~(\ref{eq:2order3})); the contribution of continuum states (second term) is negligible for the lowest states, but grows with the vibrational quantum number and becomes significant for $v=6,7$ and 8 where it amounts to 0.1, 0.2 and 0.5 kHz respectively. For the proton finite-size contributions we use a set of values where the Zemach radius is deduced from the hyperfine splitting in H~\cite{Shabaev,Karsh97,BodYen88,Fau02}, with the aim of performing a consistency check between the results of hfs spectroscopy of H and H$_2^+$. The accuracy of our new values of $b_F$ is limited by uncalculated higher-order QED corrections, which are expected to be smaller than 1~kHz.

To compare our theoretical results with those from experiment, we extract an experimental value of $b_F$ from the measured hyperfine frequencies following the procedure described in~\cite{HFS09}. The experimental accuracy claimed in Ref.~\cite{Jeff69} translates to an uncertainty margin on $b_F$ of 1~kHz. Less precise experimental values are available for the $v=0$ state~\cite{Fu92, Ost04}. The comparison with our theoretical values is presented in Table~\ref{comparison}. Theoretical and experimental results are in excellent agreement (\textit{i.e.} well within the 1-kHz experimental uncertainty). This agreement strongly supports our finding that accounting for the vibrational excitations is of essential necessity.

The work presented here is of interest for several reasons. Firstly, it successfully concludes a series of theoretical efforts, spanning nearly half a century, to fully explain the measured H$_2^+$ hyperfine structure reported in Ref.~\cite{Jeff69} within its 1-ppm uncertainty. Secondly, our results indicate that the nuclear properties of the proton as they are inferred from the hfs of atomic hydrogen are compatible with the available theoretical and experimental data on H$_2^+$. This is expected for H$_2^+$, whose electronic ground state can be written, in the lowest level of approximation, as a linear superposition of two atomic 1$s$ orbitals, distributed over the two nuclei. Thirdly, our findings suggest that for future improvements of the theoretical hyperfine structure of H$_2^+$, the $(Z \alpha)^2 E_F$ order contribution should be evaluated entirely within the three-body framework. This poses a serious challenge for theory, since cancellation of divergent parts in the three-body framework will be substantially more complicated than for the two-center adiabatic approximation.

\textbf{Acknowledgements.} This work was supported by UPMC and by the bilateral French-Dutch Van Gogh program, which is gratefully acknowledged. V.I.K. acknowledges support of the Russian Foundation for Basic Research under Grant No.~15-02-01906-a. J.C.J.K. thanks the Netherlands Technology Foundation (STW) and the Netherlands Foundation for Fundamental Research on Matter (FOM) for support. J.-Ph.K. acknowledges support as a fellow of the Institut Universitaire de France.


\begin{thebibliography}{99}

\bibitem{Jeff69} K.B.~Jefferts, Phys.\ Rev.\ Lett.\ \textbf{23}, 1476 (1969).

\bibitem{Fu92} Z.W.~Fu, E.A.~Hessels, and S.R.~Lundeen, Phys.\ Rev.~A \textbf{46}, R5313 (1992).

\bibitem{Ost04} A.~Osterwalder, A.~W\"uest, F.~Merkt, and Ch.~Jungen, J.\ Chem.\ Phys.~\textbf{121}, 11810 (2004).

\bibitem{Luke69} S.K.~Luke, Astrophys.~J.\ \textbf{156}, 761 (1969).

\bibitem{McE78} R.P.~McEachran, C.J.~Veenstra, and M.~Cohen, Chem.\ Phys.\ Lett.\ \textbf{59}, 275 (1978).

\bibitem{Babb92} J.F.~Babb and A.~Dalgarno, Phys.\ Rev.\ Lett.\ \textbf{66}, 880 (1991); J.F.~Babb and A.~Dalgarno, Phys.\ Rev.\ A \textbf{46}, R5317 (1992).

\bibitem{HFS06} V.I.~Korobov, L.~Hilico, and J.-Ph.~Karr, Phys.\ Rev.~A \textbf{74}, 040502(R) (2006).

\bibitem{HFS09} V.I.~Korobov, L.~Hilico, and J.-Ph.~Karr, Phys.\ Rev.~A \textbf{79}, 012501 (2009).

\bibitem{Oka92} T.~Oka, Rev.\ Mod.\ Phys.\ \textbf{64}, 1141 (1992).

\bibitem{Penzias68} A.A.~Penzias, K.B.~Jefferts, D.F.~Dickinson, A.E. Lilley, and H. Penfield, Astrophys.~J.\ \textbf{154}, 389 (1968).

\bibitem{Jeff70} K.B.~Jefferts, A.A.~Penzias, K.A.~Ball, D.F.~Dickinson, and A.E. Lilley, Astrophys.~J.\ \textbf{159}, L15 (1970).

\bibitem{Shuter86} W.L.H.~Shuter, D.R.W.~Williams, S.R.~Kulkarni, and C.~Heiles, Astrophys.~J.\ \textbf{306}, 255 (1986).

\bibitem{Varsh93} D.A.~Varshalovich and A.V.~Sannikov, Pis'ma Astron.\ Zh.\ \textbf{19}, 719 (1993) [Astron.\ Lett.\ \textbf{19}, 290 (1993)].

\bibitem{Pohl10} R.~Pohl {\em et al.}, Nature \textbf{466}, 213 (2010).

\bibitem{Babb08} J.F. Babb, in {\em Contributions to atomic, molecular, and optical physics, astrophysics, and atmospheric physics. Proceedings of the Dalgarno Celebratory Symposium}, ed. A. Dalgarno, J.F. Babb, K. Kirby, and H.R. Sadeghpour (Imperial College Press, London, 2010), pp. 315-319.

\bibitem{Shabaev} A.V.~Volotka, V.M.~Shabaev, G.~Plunien, and G.~Soff, Eur.\ Phys.~J.~D \textbf{33}, 23 (2005).

\bibitem{Breit} G.~Breit, Phys.\ Rev.\ \textbf{35}, 1447 (1930).

\bibitem{Kroll} N.~Kroll and F.~Pollock, Phys.\ Rev.\ \textbf{84}, 594 (1951); \emph{ibid.} \textbf{86}, 876 (1952)

\bibitem{Schwin} R.~Karplus. A.~Klein, and J.~Schwinger, Phys.\ Rev.\ \textbf{84}, 597 (1951).

\bibitem{Layzer} A.J.~Layzer, Bull.\ Am.\ Phys,\ Soc.\ \textbf{6},514 (1961); A.J.~Layzer, Nuovo Cimento \textbf{33}, 1538 (1964).

\bibitem{Zemach} A.C.~Zemach, Phys.\ Rev.\ \textbf{104}, 1771 (1956).

\bibitem{Karsh97} S.G.~Karshenboim, Phys.\ Lett.~A \textbf{225}, 97 (1997).

\bibitem{BodYen88} G.T.~Bodwin and D.R.~Yennie, Phys.\ Rev.~D \textbf{37}, 498 (1988).

\bibitem{Fau02} R.N.~Faustov and A.P.~Martynenko, Eur. Phys.~J.~C \textbf{24}, 281 (2002).

\bibitem{Carlson} C.E.~Carlson, {\em Proton Structure Corrections to Hydrogen Hyperfine Splitting}, Lect. Notes in Phys.\ \textbf{745}, 93 (Springer, 2008).

\bibitem{SapYen} J.R.~Sapirstein, D.R.~Yennie, in: T.~Kinoshita (Ed.), {\em Quantum Electrodynamics}, World Scientific, Singapore, 1990.

\bibitem{Kin96} T.~Kinoshita and M.~Nio, Phys.\ Rev.~D \textbf{53}, 4909 (1996).

\bibitem{var00} V.I.~Korobov, Phys.\ Rev.~A \textbf{61}, 064503 (2000).







\end{thebibliography}
\end{document}